\documentstyle[aps,preprint]{revtex}
\begin{document}
\title{Effect of Hydrostatic Pressure on the Kondo Effect in the Heavy Fermion System  Yb$_2$Fe$_3$Si$_5$}
\author{Yogesh Singh and S. Ramakrishnan}
\address{Tata Institute of Fundamental Research, Mumbai-400005, India}
\author{C. Geibel}
\address{Max-Planck-Institute for the Chemical Physics of Solids, D-01187 Dresden, Germany}
\maketitle
\begin{abstract}
\noindent 
We report the effect of hydrostatic pressure on the resistivity of the heavy fermion Kondo lattice compound Yb$_2$Fe$_3$Si$_5$ in the temperature range 2-300~K and in pressures upto 10 Kbar. At ambient pressure the resistivity shows a prominent Kondo effect. The Kondo minimum shifts from 22.5~K at ambient pressure to 19.2~K at a pressure of 10~Kbar. The depth of the Kondo minimum also progressively decreases with pressure. These results suggest a possible weakening of the Kondo effect with the application of pressure.    
\vskip 1truecm 
\noindent     
Ms number ~~~~~~~~~~~~PACS number:~72.10.Fk, 72.15.Qm, 75.20.Hr, 75.30.Mb\\
\end{abstract}

\newpage
\section{Introduction}
\label{sec:INTRO}
\noindent
Compounds of the series R$_2$Fe$_3$Si$_5$ are known to show unusual superconducting and magnetic properties \cite{r1,r2,r3,r4}. Both Tm$_2$Fe$_3$Si$_5$ and Lu$_2$Fe$_3$Si$_5$ of this series undergo superconducting transitions at low temperatures. While Lu$_2$Fe$_3$Si$_5$ has the highest T$_C$ ( =~6~K) for an Fe containing compound, Tm$_2$Fe$_3$Si$_5$ is the first reentrant superconductor where the superconductivity at about 1.6~K is destroyed by the onset of antiferromagnetic order at 1~K.\\
Given that both the Tm and Lu samples of the series show superconductivity at low temperatures, the absence of superconductivity in Yb$_2$Fe$_3$Si$_5$ is puzzling. Our recent work \cite{r5} on high quality single-phase polycrystalline samples of Yb$_2$Fe$_3$Si$_5$ revealed the presence of strong hybridization ($\gamma$=~500~mJ/Yb mol K2 ) and Kondo effect below 25 K in this sample. It is known that Kondo interactions can strongly suppress superconductivity. For Ce compounds, pressure is known to enhance the Kondo temperature T$_K$. Yb being the hole analog of Ce (Ce has one 4f electron while Yb has one 4f hole), the T$_K$ for Yb systems can be expected to go down on the application of pressure.
Thus, we have studied the effect of the application of hydrostatic pressure on the Kondo effect. Our aim is to understand the absence and possibility of inducing superconductivity in this compound by the application of pressure.  
 
\section{EXPERIMENTAL DETAILS}
\label{sec:EXPT}
\noindent
Polycrystalline samples of Yb$_2$Fe$_3$Si$_5$ were prepared by the closed crucible method described in our earlier work \cite{r5}. Powder X-ray diffraction measurements confirmed the structure and the absence of impurity phases.\\
The dc resistivity measurements were performed on a bar shaped sample of Yb$_2$Fe$_3$Si$_5$ 
using contacts made with silver paste. The sample is placed inside a teflon tube filled with
a hydrostatic pressure medium (Daphne oil). This teflon tude is kept inside a
Cu-Be cell of the piston-clamp type which can be pressurised upto 14 Kbar. The resistivity was measured using an ac resistance bridge (Linear Research Inc., USA) from 2 to 300~K.

\section{RESULTS AND DISCUSSION}
\label{sec:RES}
\noindent
In fig~1 we show the ambient pressure resistivity $\rho$(T) for Yb$_2$Fe$_3$Si$_5$ in the temperature range 2-300~K. The resistance decreases slowly as we cool down from 300~K untill about 150~K after which it starts to drop more rapidly indicative of crystal field effects.
The inset~(a) shows the resitivity between 0.4 and 2.5~K. A distinct change of slope at 1.7~K marks the onset of the anti-ferromagnetic order in the sample. This is consistent with the resistivity, susceptibility and heat capacity data on our earlier sample \cite{r5}.\\ 
The inset~(b) in fig.~1 shows the data between 2-50~K on an expanded scale to emphasize the presence of the Kondo effect with a minimum in resistivity at 22.5~K after which the resistance starts to increase again until it reaches a maximum at around 6~K after which it sharply decreases again due to coherent scattering. We will call the maximum in the resistivity as the coherence peak and the temperature at which it occurs as the coherence temperature T$_{coh}$ to make it easy to compare the data for different pressures.\\
The fig.~2 shows the resistivity of Yb$_2$Fe$_3$Si$_5$ for various pressures. We have plotted R(T)/R(300~K) to be able to make qualitative comparisons between the data for different pressures. The panel~I shows the normalized resistance between 2-300~K for applied pressure values P = 0, 3.5, 5, 8.5 and 10~Kbar. P= 0~Kbar actually means the ambient pressure data. At present the values of the pressures are taken at room temperature. It must be noted that the actual pressure at low temperatures will be slightly reduced. From panel~I it is evident that the data for all pressures fall on top of each other between 100- 300~K. It can be safely said that the pressure has no appreciable effect on the lattice contribution to the transport properties. The data for different pressures begin to deviate only below 50~K where the Kondo effect is dominant. 
\par
In panel~II we show the data between 2-50~K on an expanded scale to highlight the effect of pressure on the Kondo effect. It can be seen that the Kondo minimum shifts to lower temperatures on the application of pressure and the depth of the Kondo minimum also progressively decreases with increasing pressure.\\
To make a quantitative comparison we have listed in table~1 the values of the applied pressure P, T$_{MIN}$ at which the Kondo minimum occurs, T$_{COH}$ at which the coherence peak occurs and R$_{TCOH}$/R$_{TMIN}$ as a measure of the depth of the Kondo minimum.\\
It can be seen from the second column listing the values of T$_{MIN}$ that as we increase the applied pressure, the Kondo minimum shifts to lower temperatures. Also, from the fourth column in which we list values of R$_{TCOH}$/R$_{TMIN}$ as an estimate of the depth of the Kondo bowl, it is evident that the Kondo bowl becomes progressively shallower on increasing pressure.
There is no appreciable change in the temperature T$_{COH}$ of the onset of the coherence peak. 
    
\section{CONCLUSION}
\label{sec:CON}
We have studied the pressure dependence of the resistivity of the heavy fermion Kondo lattice compound Yb$_2$Fe$_3$Si$_5$ to look at the effect of hydrostatic pressure on the Kondo interactions in this compound and in Yb compounds in general.\\
We find that the temperature T$_{MIN}$ at which the Kondo minimum occurs, reduces from about 22.5~K at ambient pressure to 19.2~K at a pressure of 10~Kbar. We have also seen a monotonic reduction of the depth of the Kondo minimum with increasing pressure. Taken together we can safely say that the application of pressure has the effect of weakening the Kondo effect in the compound Yb$_2$Fe$_3$Si$_5$. Further studies are in progress to see whether we can induce superconductivity in this compound via the suppression of the Kondo effect with the application of pressure.

\begin{figure}
\caption{Temperature dependence  of the resistivity ($\rho$)
of Yb$_2$Fe$_3$Si$_5$ from 0.4 to 300~K. The inset (a) shows
the $\rho$ data from 0.4 to 2.5~K on an expanded scale. A definite 
change of slope below 1.7~K marks the onset of the anti-ferromagnetic ordering. The solid line is the fit to a T$^2$ law.
The inset (b) shows $\rho$ data upto 50~K. A distinct Kondo effect is visible with a minimum in $\rho$ at 22.5~K and a drop below 6~K due to coherence. 
\label{fres1}}
\end{figure}
\begin{figure}
\caption{Temperature dependence of the resistivity at various pressures, normalized to it's room temperature value. The panel~I shows the data between 2-300~K. It is clear that there is no significant effect of pressure on the lattice part of the resistivity. The panel~II shows the data between 2-50~K on an expanded scale to show more clearly  the effect of pressure on the Kondo effect.
\label{fres2}}
\end{figure}

\newpage
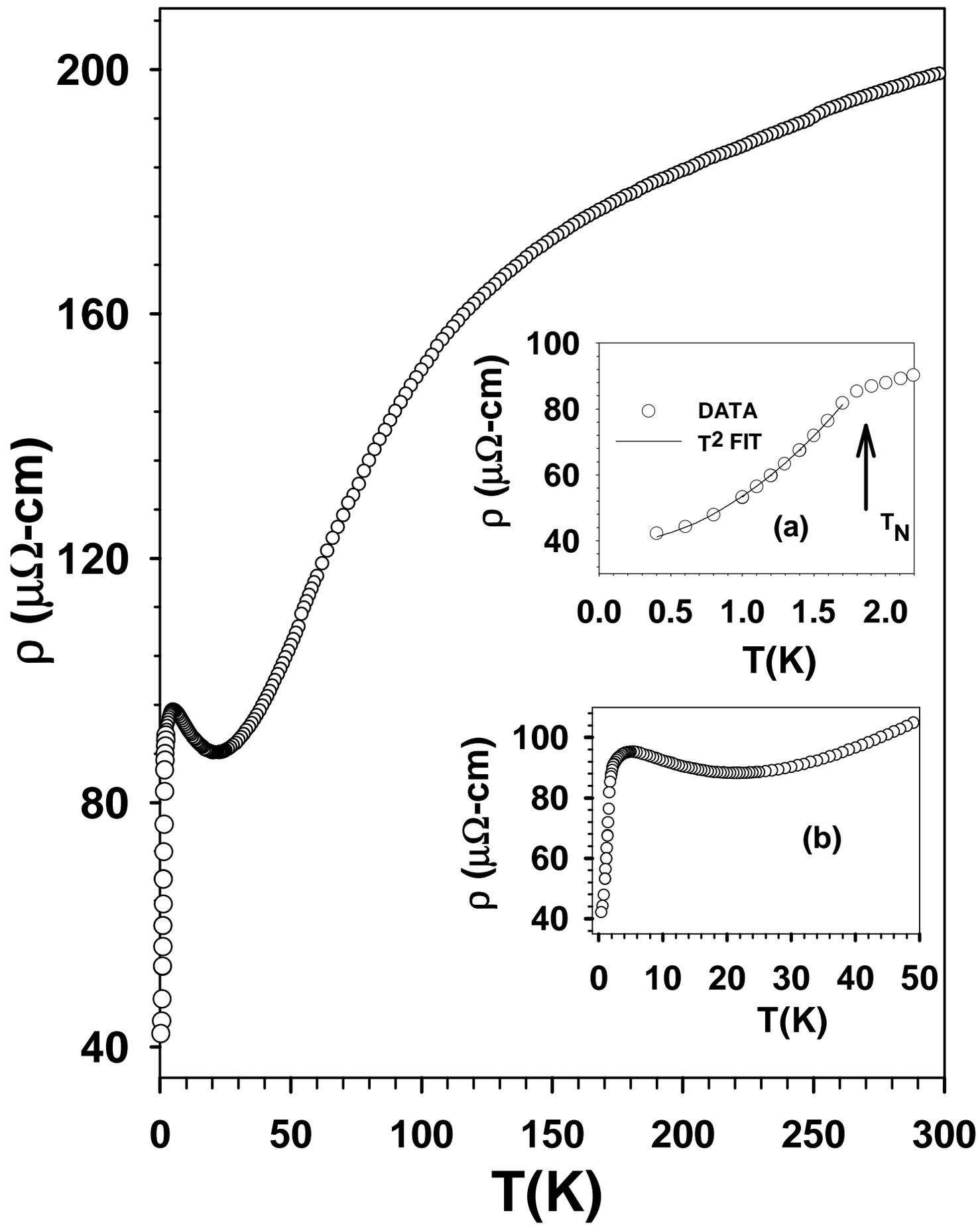
\begin{table}
\caption{Parameters obtained from the resistivity data of Yb$_2$Fe$_3$Si$_5$ under pressure.}
\begin{tabular}{ccccc}
P(Kbar)&~T$_{MIN}$ & ~T$_{COH}$ & ~R$_{TCOH}$/R$_{TMIN}$ &\\
 \tableline
0&22.5&4.68&1.063&\\
3.5&21.3&4.74&1.059&\\
5&20.8&4.68&1.056&\\
8.5&20.1&4.72&1.053&\\
10&19.2&4.75&1.049&\\
\end{tabular}
\label{table 1}
\end{table}
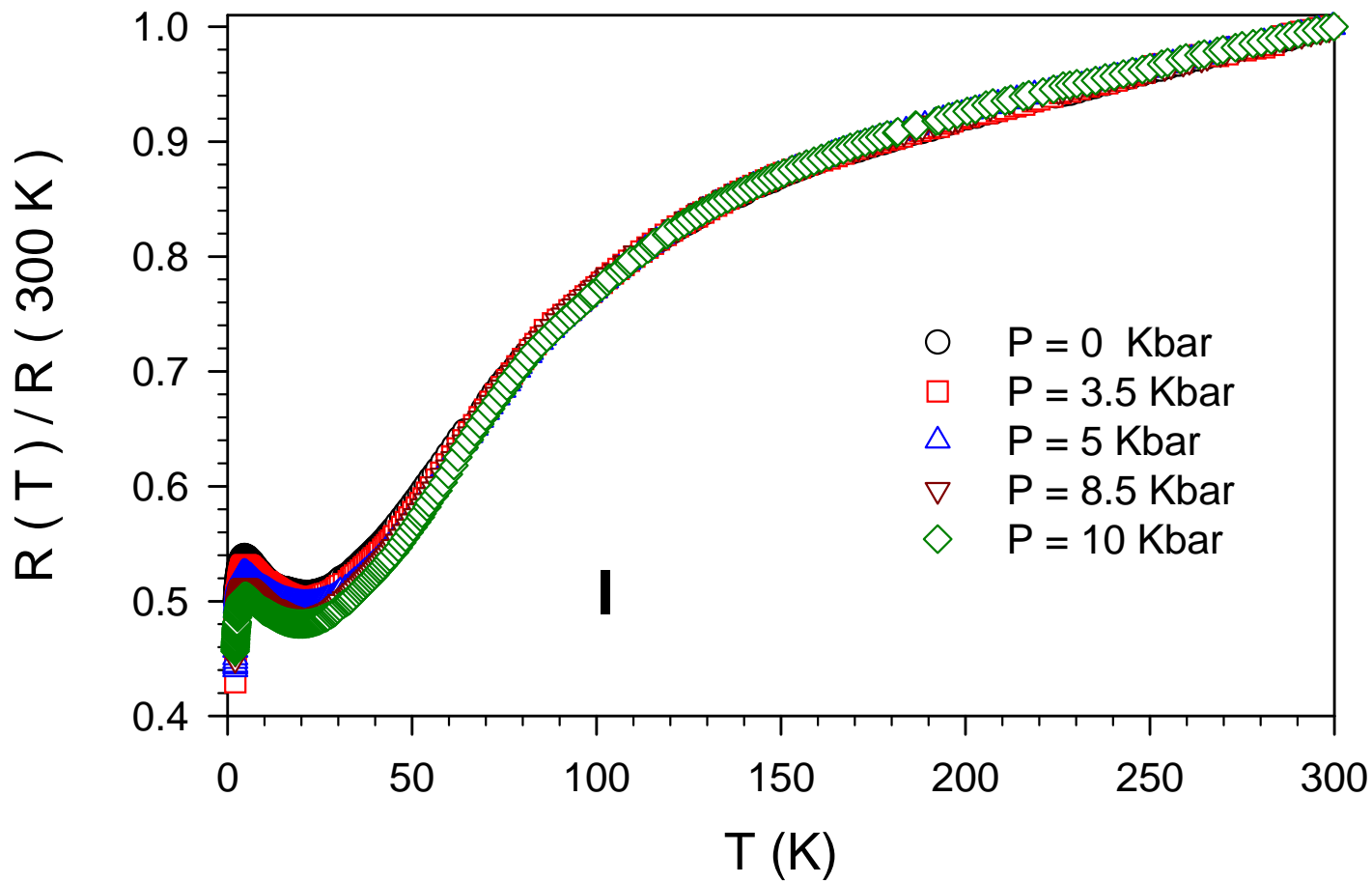
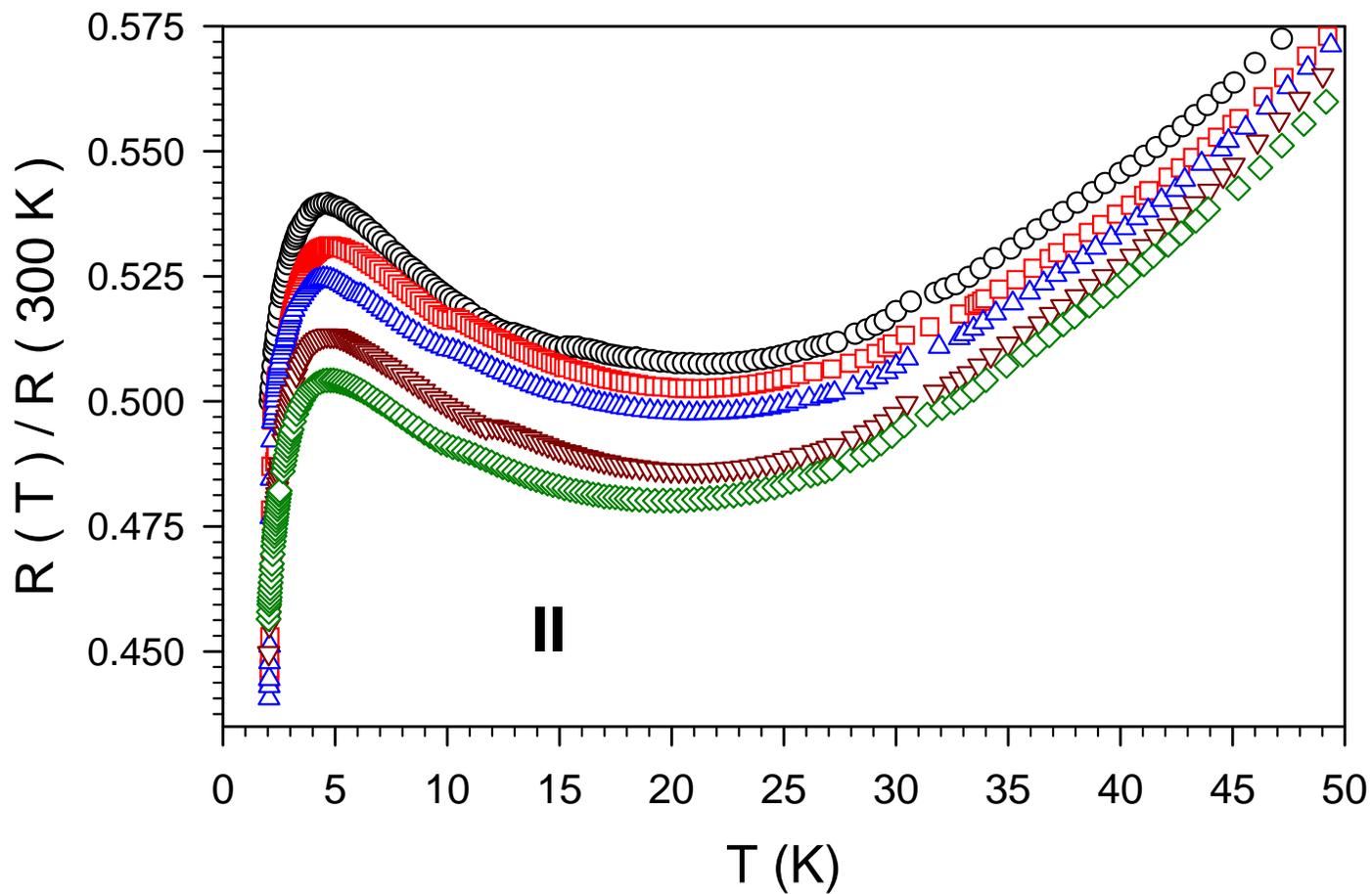
\end{document}